\documentclass[copyright,creativecommons]{eptcs}
\providecommand{\event}{DCDP 2010} 
\usepackage{breakurl}             
\usepackage[pdftex]{graphicx}
\title{Separating Agent-Functioning and Inter-Agent Coordination by Activated Modules: The DECOMAS Architecture}
\author{Jan Sudeikat\footnote{Jan Sudeikat is doctoral candidate at the Distributed Systems and Information Systems (VSIS) group, Department of Informatics, Faculty of Mathematics, Informatics and Natural Sciences, University of Hamburg, Vogt--K\"olln--Str. 30, 22527 Hamburg, Germany, jan.sudeikat@informatik.uni-hamburg.de}  and Wolfgang Renz
\institute{Multimedia Systems Laboratory (MMLab),\\
Faculty of Engineering and Computer Science, Hamburg University of Applied Sciences,\\
Berliner Tor 7, 20099 Hamburg, Germany}
\email{\textbraceleft jan.sudeikat|wolfgang.renz\textbraceright @.haw-hamburg.de}
}
\def\titlerunning{Separating Agent-Functioning and Inter-Agent Coordination by Activated Modules}
\def\authorrunning{J. Sudeikat \& W. Renz}
\begin{document}
\maketitle

\begin{abstract}
The embedding of \emph{self-organizing} inter-agent processes in distributed software applications enables the decentralized coordination system elements, solely based on concerted, localized interactions. The separation and encapsulation of the activities that are conceptually related to the coordination, is a crucial concern for systematic development practices in order to prepare the reuse and systematic integration of coordination processes in software systems.  
Here, we discuss a programming model that is based on the externalization of processes prescriptions and their embedding in \emph{Multi-Agent Systems} (MAS). One fundamental design concern for a corresponding execution middleware is the minimal-invasive augmentation of the activities that affect  coordination. This design challenge is approached by the \emph{activation} of agent modules. Modules are converted to software elements that reason about and modify their host agent. We discuss and formalize this extension within the context of a generic coordination architecture and exemplify the proposed programming model with the decentralized management of (web) service infrastructures.
\end{abstract}
\section{Introduction}
\emph{Self-Organization} describes adaptive processes among system elements, as found in physical, biological, and social systems, that establish and maintain structures \cite{Prokopenko2008}. The utilization of self-organization principles is an alternative approach for the construction of self-adaptive, distributed software systems \cite{Salehie2009}. This approach is attractive, as it allows to embed adaptive properties in the interplay of system entities. Consequently, centralized responsibilities are avoided that may imply bottle necks and single points of failure. 
Self-organizing system phenomena are governed by feedback loops, i.e. circular interdependencies among system elements (e.g. discussed in \cite{Bonabeau1999}). Unlike the control loops in self-managing software systems \cite{Brun2009}, the loops are decentralized, i.e. distributed among system elements.

In the research project "Selbstorganisation durch Dezentrale Koordination in Verteilten Systemen"\footnote{Self-Organisation by Decentralized Coordination in Distributed Systems} (Sodeko VS), the utilization of self-organizing inter-agent processes as reusable design elements is studied \cite{Sudeikat2009d}. Distributed feedbacks, as structures of mutual influences among system elements, are elevated to discrete design elements. These structures are used to define inter-agent processes and a corresponding programming model can be used to integrate these processes in agent-based software systems. A foundational building block is a middleware layer that provides an execution context for the process enactment and integration \cite{Sudeikat2009d} (see Section \ref{sect:architecture}).


A key design criterion is that adaptive features can be supplemented to functioning sets of software agents, i.e. \emph{Multiagent Systems} (MAS). Developers can add decentralized coordination, i.e. the self-organization of system aspects, to working systems. A prerequisite is the conceptual and practical separation of the activities that are conceptually related to the inter-agent coordination. These activities concern the participation in a coordinating process and define a supplement, which influences the core functionality of the agents. In this paper, we present an approach for this separation that is based on extending agent-oriented implementation modules. This allows the minimal-intrusive encapsulation and automation of inter-agent coordination. In addition, the discussed enhancement is attractive for MAS developers as it allows to modularize crosscutting concerns in MAS (see Section \ref{sect:crosscutting}).


This paper is structured as follows. In the following section, related work is outlined. In Section \ref{sect:architecture}, a programming model for self-organization is outlined. The technological foundation for the encapsulation and automation of coordination activities is the activation of agent modules that is discussed in Section \ref{cc-in-aose}. 
The utilization of the programming model is exemplified in Section \ref{sect:shoaling_glassfishes} before we conclude and give prospects for future work (see Section \ref{sect:conclusions}).
\section{Related Work}\label{sect:related-work}
Agent technology provides tools and concepts for the construction of autonomous software elements and is a prominent grounding for the development of self-organizing applications \cite{Serugendo2006}. Natural self-organizing systems are composed of \emph{autonomous} system elements \cite{Prokopenko2008}, e.g. particles and cells, and the coaction of these elements can be metaphorically resembled with autonomous software agents.   
\subsection{Integrating Coordination Mechanisms}\label{sect:mechanisms}
The construction of self-organization, i.e. an adaptive, coordinating process that structures the configurations of system elements, is based on two foundational types of implementation mechanisms \cite{Sudeikat2007}. First, generic \emph{interaction}-level mechanisms have been proposed that allow to establish information flows between system elements (e.g. reviewed in \cite{Wolf2006a}). Among others, these mechanisms support the stochastic dissemination of information and the attenuation of outdated data. Secondly, adaptation-level mechanisms control the participation in interactions, the processing of the exchanged information, and affect the conclusive adjustments within software agents that result from the exchanged information.  

The encapsulation of interaction-level mechanisms is typically approached by dedicated communication infrastructures and languages \cite{Gelernter1992}. These are means to decouple software components but the coordination logic, e.g. when to interact and how to (locally) respond to interactions is blended in the control flow the system elements. In previous works, three foundational approaches have been followed to separate this control of the coordination from the control of the element functioning. First, specialized agent architectures (e.g. \cite{Shabtay2006}) have been proposed that outsource coordination-related activities to specific \emph{modules}. Secondly, the separation can be enforced by the \emph{execution infrastructure} that is given by the utilized programming-language or middleware. An example is the outsourcing of coordination by using aspect-orientation \cite{Seiter2006}. Finally, approaches use networked elements to control the localized adjustments \cite{Serugendo2009}.

In this paper, a coordination middleware layer is presented that extends the modularization-based approaches (see above) to enable the separation and integration of coordination logic in \emph{generic} agent architectures. Preparing the integration in established, general-purpose agent-architectures allows to reuse the existing constructive knowledge and tool support that concerns these agent models, e.g. methodology-specific agent design techniques. The direct integration, by reusing agent-modularization concepts avoids the communication overhead of externalized approaches.     
\subsection{Crosscutting Concerns in Agent-Orientation}\label{sect:crosscutting}
Modularization enforces the decomposition of software systems into functional clusters with minimum overlapping functionality. These so-called \emph{core concerns} are typically separated into different components or modules \cite{Parnas1972}, complex  software systems often comprise additional \emph{crosscutting concerns}, so-called \emph{aspects} \cite{Kiczales1997}, which are to be referenced from various modules. Prime examples are amongst others \emph{failure recovery}, \emph{monitoring} and \emph{logging}. While these functionalities can be clustered in modules, these will be referenced throughout the agent model. Thus the information when to invoke the functionality is spread and references are scattered. In this respect, the embedding of coordination is regarded as another crosscutting concern. When the logic how to adjust and interact, as to participate in a collaborative process, is encapsulated in specific module, the contained activities have to be frequently referenced in the agent model and these references with be scattered as well. Consequently, it is desirable to contains the functionality, as well as the context if it's invocation in a single agent element.  

The notion of \emph{crosscutting} concerns for agent modularization has to date found minor attention. 
Numerous MAS infrastructures are build with object-oriented programming languages, thus Aspect-oriented Programming (AOP) \cite{Kiczales1997} is one approach to embed crosscutting functionalities with additional programming language tool sets. 
In \cite{Lobato2004}, it is exemplified how AOP techniques can be used to modularize object-oriented agent models by encapsulating mobility related API calls that are available in the JADE\footnote{http://jade.tilab.com/} agent platform and Garcia et al. \cite{Garcia2005a}, examined how AOP frameworks facilitate the realization of object-oriented agents. Examples of aspects in software agents are \emph{interaction}, \emph{mobility} and \emph{learning} \cite{Garcia2004}. 
\subsection{Agent Modularization Using the Example of BDI Agents}\label{bdi-se}

Agent platforms \cite{Bordini2006} provide distributed middlewares for the construction of MAS. One prominent architectural model is the \emph{Belief Desire Intention} (BDI) architecture \cite{Rao1995} that allows to express both longterm goal-directed objectives as well as reactivity. Following this architectural style, agents are structured as sets of \emph{Beliefs}, \emph{Goals}, and \emph{Plans}. Beliefs contains the local knowledge of the agent about itself and the environment. Goals represent the objectives and plans are the executable means of agents. BDI-specific reasoning engines control the agent execution. The currently active goals are deliberated and means-end reasoning is used to select plans for the achievement of goals.     
Modularity in terms of functional independent clusters has been introduced to BDI agents by the \emph{Capability} concept \cite{Busetta2000,Braubach2005}. Capabilities describe clusters of BDI concepts, i. e. Beliefs, Goals and Plans in a name-space. These enable the recursive inclusion of other capabilities (sub-capability). The interplay with a surrounding agent/capability (super-capability) is controlled by scoping rules. The visibility of the comprised elements is specified as well as the visibility of relevant events, generated outside of the capability. 

In \cite{Riemsdijk2006}, a goal-centric modularization scheme, so-called \emph{Goal-Oriented Modularity}, has been proposed. Modules encapsulate the information how sets of related goals can be satisfied. This enables a higher degree of encapsulation of behaviors.
A behavior-based stance towards agent modularization is given in \cite{Dastani2005} where \emph{role} concepts encapsulate sets of beliefs, goals, plans and reasoning rules. In addition, the \emph{Enactment} and \emph{deactment} of roles at run-time is prepared. Modularization by \emph{policy-based intentions} is proposed in \cite{Hindriks2008}. Developers can explicitly declare in which context a module is to be activated. 

Due to the widespread, recognition, and practicability of the BDI agent model, the prototype realization of the here discussed coordination middleware (see Section \ref{sect:architecture}) is based on this agent type. In this paper, the utilized enhancements to modularize \emph{crosscutting concerns} are discussed in general (see Section \ref{cc-in-aose}) and are then detailed for this particular agent model (see Section \ref{sect:cc-bdi}).
\section{The DECOMAS Architecture}\label{sect:architecture}
Within the SodekoVS research project \cite{Sudeikat2009d}, a programming model for distributed feedbacks among system elements is revised. 
A key objective of this framework is that the ability to self-organize is provided as an optional tool that development teams can integrate in their applications when needed. The aim is to enable the supplementation of self-organizing properties when the need for decentralized coordination of system elements is revealed.
The foundational elements are a declarative configuration language \cite{Sudeikat2009} and an architectural model for the supplementation of externally prescribed inter-agent processes. Here, the configuration of self-organizing processes is not discussed. Details on the configuration language can be found in \cite{Sudeikat2009} and a graphical representation of the process description is exemplified in the Sections \ref{sect:integrating_servier_utilization_management} and \ref{sect:integrating_service_balancing} to illustrate the intended application dynamics.

\begin{figure}[htbp]
\center
\includegraphics[width=0.8\textwidth]{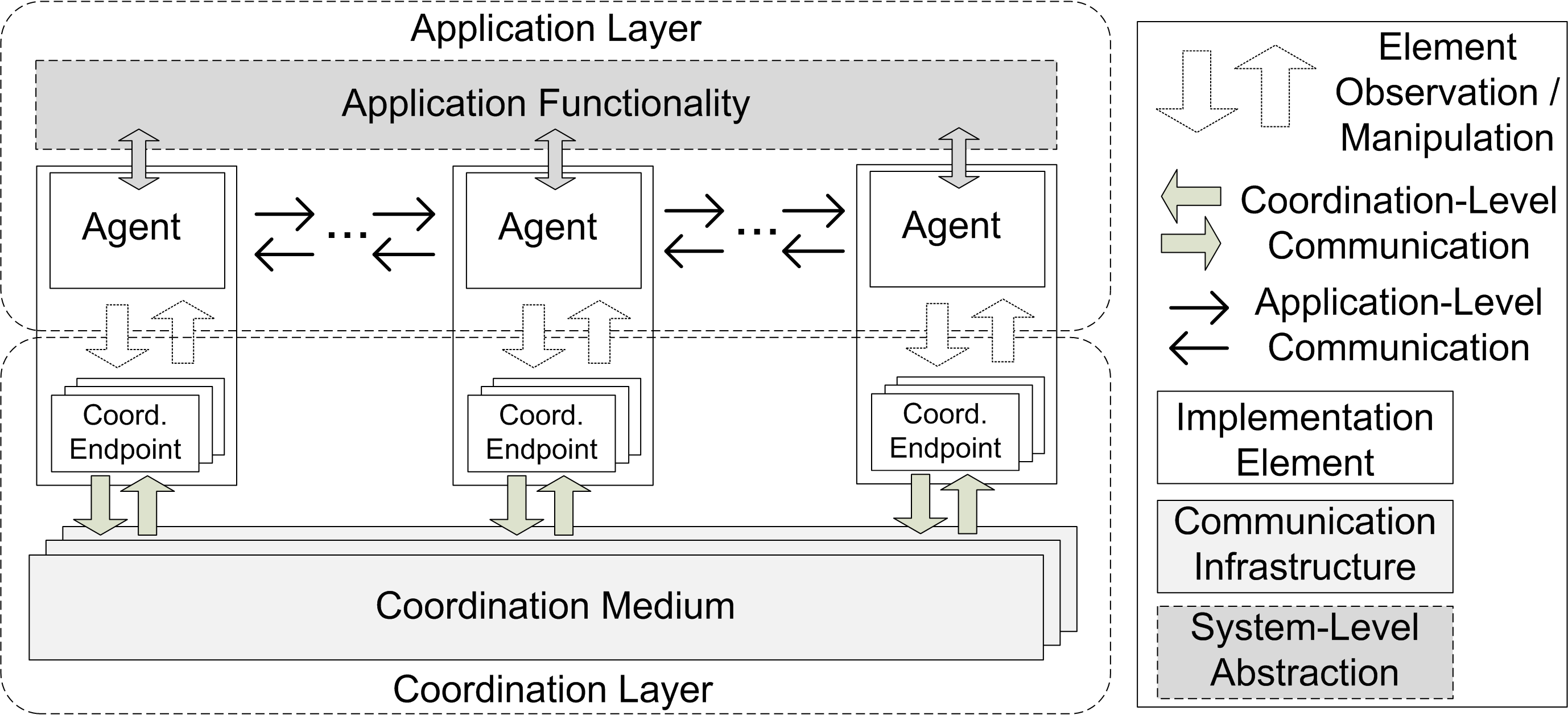}
\caption{The SodekoVS-Architecture to the embedding of decentralized coordination in MAS \cite{Sudeikat2009c}.}
\label{fig:sodeko-architecture}
\end{figure}

This integration architecture follows a layered structure that is illustrated in Figure \ref{fig:sodeko-architecture}. The \emph{Application Layer} contains the application functionality. Within this layer, agents act as providers of application-dependent functionalities. An underlying \emph{Coordination Layer} controls  the enactment of coordinating processes among (sub-)sets of agents. \emph{Coordination Media} are conceptual entities that contain interaction mechanisms \cite{Wolf2006a}. Inside these media, these are realized by the utilization of communication infrastructures (e.g. see \cite{Gelernter1992}). The details of the interaction mechanisms are hidden by a generic publish/subscribe interface. 
The utilization of these media is shielded from the agent internals by intermediate \emph{Coordination Endpoints}. These are associated to an agent and control the participation in a coordination process. Endpoints are enabled to observe and modify the execution of associated agents. The rationale is that Endpoints interact, via Media, in place of the associated agent and decide the local adjustments. Therefore, coordination-relevant activities, including adaptation-level mechanisms, are encapsulated (see Section \ref{sect:mechanisms}). The operation of Endpoints are declaratively configured (see above). These declarations indicate which changes in the agent-internal configuration are significant for their participation in an inter-agent process. These events are then propagated via Media and processed by perceiving Endpoints. If these perceptions indicate the need for adjustments, these are made by triggering agent-internal behaviors. Agent models are often capable to show concurrent conduct of behaviors and their scheduling is realized in the agent execution environment (e.g. see Section \ref{bdi-so}). Agents can be associated to more then one Endpoint, so the enactment of different processes is separated as well. A prototypical implementation of this architecture is reported in \cite{Sudeikat2009j} and it's utilization is exemplified in \cite{Sudeikat2009c}. 

\section{Activating Agent Modules: Enabling Contributive Processing}\label{cc-in-aose}
Agent-oriented software development is supported by comprehensive development environments. These provide execution middleware and programming languages for the utilization of agent-based implementation concepts \cite{Bordini2006}. It is desirable that agent developers can utilize these concepts throughout the whole development cycle, also when expressing cross-cutting concerns (see Section \ref{sect:crosscutting}). Conventional modules cluster functional concerns. These are typically used inside agents by explicitly referencing contained elements, e.g. dispatching (sub-)goals that are contained modules \cite{Braubach2005}.
The aim of the proposed extension is to automate these references. Both the functionality and the information when it is to be invoked are contained in modules.
These modules extend conventional agent modules, as modules are equipped with the ability of observe and modify the agent execution. These enhanced modules operate as autonomous actors that react to changes in the immediate context, i.e. the state of their host agent or super module. 
We name these modules \emph{co-efficient}, since they register for contributive processing on certain agent reasoning events. The presented module concepts allows to compose agents as sets of independent actors. 
Besides the structuring of agent models, these modules facilitate the embedding of crosscutting mechanisms, like logging, failure recovery etc., to be automatically triggered. A example is the encapsulation of the  monitoring of agent-behaviors in \cite{Sudeikat2006}. A module observes the reasoning of the host agent and decides the recording, when the course of action is significantly adjusted. 

\begin{figure}[htbp]
\center
\includegraphics[width=0.9\textwidth]{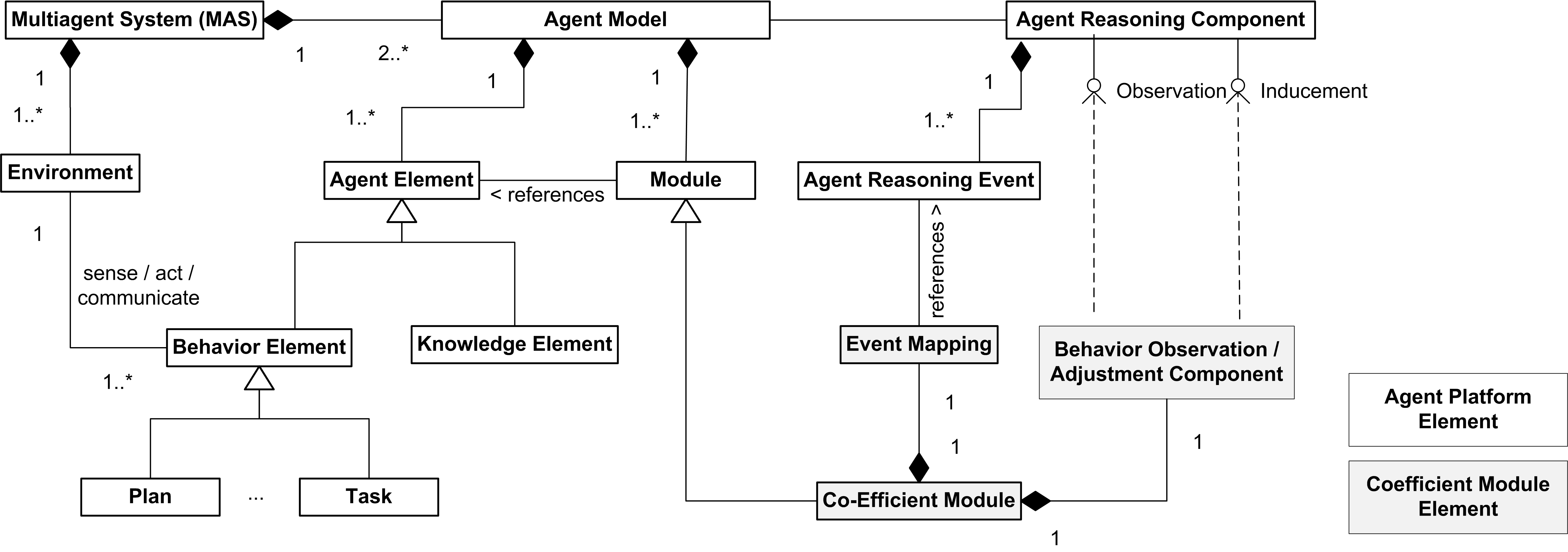}
\caption{Conceptual model of co-efficient agent modules.}
\label{fig:meta-model}
\end{figure}

Figure \ref{fig:meta-model} illustrates the conceptual model of a co-efficient agent module. Abstracting from specific agent architectures, we assume that the execution of an \emph{Agent Model} is managed by a \emph{Reasoner}, e.g. using reactive or deliberative mechanisms. The reasoning component processes \emph{Agent Reasoning Event}s that characterize the agent execution and reference agent elements which are modified. Examples are changes in agent-intern data structures (knowledge) or the execution of plans. A module concept allows to structure agents by containing sets of agent elements (e.g. \cite{Busetta2000}). Co-efficient modules extend these with two additional components. First, an \emph{Observation / Adjustment Component} allows to observe and modify agent execution by registering for and dispatching reasoning events. This component makes use of platform-specific interfaces (\emph{Observation} / \emph{Inducement}). Secondly, developers specify an \emph{Event Mapping} that describes which events are subject of observation as well as which events are to be dispatched to the reasoning mechanism.

\begin{figure}[htbp]
\center
\includegraphics[width=0.7\textwidth]{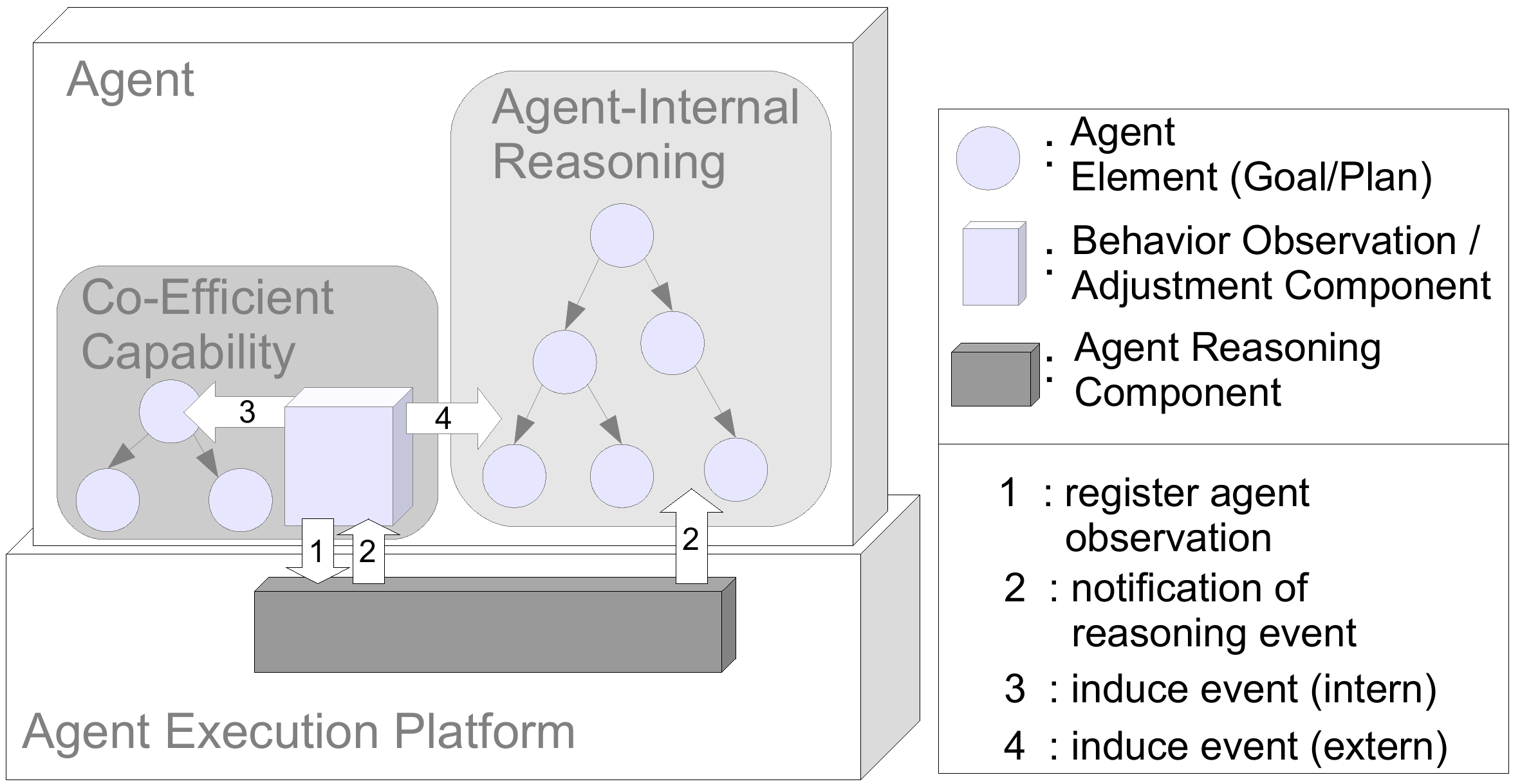}
\caption{Execution of a Co--Efficient Capability.}
\label{fig:aose06-mechanism}
\end{figure}

Figure \ref{fig:aose06-mechanism} outlines the operating principle of a coefficient module. On agent start-up the module is registering (\emph{Observation} interface) for the observation of a set of reasoning events in the surrounding agent (1). Subsequently, it is notified about these events (2) and the event mapping is interpreted to infer which events to dispatch via the \emph{Inducement} interface. These events reference agent elements (e.g. goals, beliefs) inside the module (3), or in the surrounding agent (4). The available reasoning events can be classified according to their effects on the agent execution. Events denote modifications of the agent state, the agent behavior, the agent model, or describe communicative activities. The concrete realization of these event categories depend on the utilized agent platform and architecture (e.g. see section \ref{impl}).
\subsection{Formalization of Coefficiency in Operational Semantics}\label{op-sem}\label{sect:cc-bdi}
In \cite{Rao1995} the BDI architecture has been defined by an abstract interpreter and a corresponding proof theory. This seminal work bridges the gap between the theory and practice of BDI agents and led to the logic--based \emph{Agentspeak(L)} programming language \cite{Rao1996}. Based on this language and further formalizations, the Operational Semantics for BDI agents have been given in \cite{Vieira2007} to support implementation and verification of BDI-based MAS. Here, we adopt these semantics to formally express the impact of co-efficient modules. Within BDI agents these are Coefficient Capabilities (CECs) (see Section \ref{bdi-se})).
\subsubsection{Operational Semantics for BDI Agents}\label{bdi-so}
Operational Semantics are an established formalism to describe the semantics of programming languages. The  operation of programs is expressed by a transition relation between program configurations \cite{Plotkin2004}. The complete specification of the Operational Semantics of BDI reasoning is given in \cite{Vieira2007}. In this section, key definitions are summarized that are used to specify the effects of crosscutting concerns on the execution of BDI agents (cf. Section \ref{cc-so-detail}).

An agent configuration is defined as a a tuple \(<ag,C,M,T,s>\) \cite{Vieira2007}. \(ag\) is the agent program, given by a set of \emph{beliefs} and \emph{plans}. \(C\) is a \emph{Circumstance} that resembles the execution context of an individual agent. \(M\) is a tuple that characterizes the agent communication. \(T\) is a tuple that provides temporary information that is used by the reasoner and \(s\) is the current step in the reasoning cycle.

In the following, these elements are detailed. The Circumstance $C$ is defined as a tuple \(<I,E,A>\), where the element \(I\) is a set of \emph{intentions} \(\{i,i',\ldots\}\). An intention \(i\) is a stack of partially instantiated plans. \(E\) is a set of \emph{events} \(\{(te,i),(te',i'),\ldots\}\). These are denoted as pairs \((te,i)\) of a triggering event (\(te\)) and a related intention (\(i\)). When events result from the processing of other events, e.g. from the achievement of (sub)goals, these follow-up events are associated to the currently active intention. An empty intention is denoted by $\top$. In this respect, intentions represent different courses of actions. Their concurrent execution is controlled by the agent reasoner \cite{Vieira2007}. \(A\) is a set of actions that is available to the agent to modify the environment.


The asynchronous communication of agents is characterized by the tuple \(M\). The communicative abilities of agents are not influenced by CECs, therefore the management of communications is not discussed. Details can be found in \cite{Vieira2007}.

Temporary information is kept in the tuple \(T\). The elements \(<R,Ap,\iota,\rho,\epsilon>\) provide volatile data to the reasoner. The set $R$ is the set of plans that are \emph{relevant} for the current event. These plans are capable to handle the event. $Ap$ denotes the set of plans that are not only relevant but can be activated. The elements $\iota,\epsilon,\rho$ refer to the current intention, event and applicable plan that are considered during one reasoning cycle.

Finally, the current step in the reasoning cycle is given by the element $s$. Altogether the reasoning cycle is composed of nine steps \(s \in \{ProcMsg, SelEv, RelPl, ApplPl, SelAppl, AddIm, SelInt, ExecInt,\\ClrInt \}\). These steps are: processing incoming messages, selecting an event to be handed, computing the relevant plans, computing the applicable plans, adding means, i.e. plans, to an intention, selecting an intention, executing an intention, clearing an intention \cite{Vieira2007}.

CECs affect the selection of the handled events. Therefore, we summarize here the semantics of the original \emph{Event Selection} rule (\emph{SelEv}). This rule picks an event and marks it for further processing. BDI agents employ \emph{reactive planning} they handle the events in \(E\). Events are added to \(E\) by transition rules or elements of the general architecture outside the agent interpreter, e. g. belief updates. The rule SelEv1 refers to the \emph{selection function} $S_E$ that selects events from the set \(E\). Selected events are removed from $E$ and added to $\epsilon$ for further processing. If no event is to be handled, the rule SelEv2 skips directly to the intention execution that is initialized by the selection of an intention (SelInt) \cite{Vieira2007}:  
\begin{equation}
	\textbf{SelEv1}\quad \frac{S_{E}(C_{E}) = \langle te,i\rangle }{\langle ag,C,M,T,SelEv\rangle \longrightarrow \langle ag,C',M,T',RelPl\rangle}
\end{equation}
\begin{displaymath}	
	where: C'_{E} = C_{E} \backslash \{\langle te,i \rangle\}
\end{displaymath}
\begin{displaymath}	 
	T'_{\epsilon} = \langle te,i\rangle	
\end{displaymath}

\begin{equation}
	\textbf{SelEv2}\quad \frac{S_{E}(C_{E}) = \{ \}}{\langle ag,C,M,T,SelEv\rangle \longrightarrow \langle ag,C,M,T,SelInt\rangle }
\end{equation}

In order to handle selected events, the relevant and applicable plans are calculated and one applicable plan is selected. A group of rules is responsible to execute this applicable plan. Additional rules control the execution of different intentions. \emph{External} events, i. e. events that are perceived and not generated by previous plan executions trigger the creation of a novel intention. These stacks of partially instantiated plans are added, removed, and selected for execution by dedicated transition rules \cite{Vieira2007}.
\subsubsection{The Operational Semantics for Co-Efficient Capabilities}\label{cc-so-detail}
Co--efficient modules register themselves for agent observation (see Section \ref{cc-in-aose}). These modules are notified when an event in a subset of reasoning events occurs. Upon these occurrences additional BDI reasoning events are dispatched in the surrounding agent. Realizations of these modules contain the configuration of (1) the events that are to be added by certain observations and (2) the execution context that permits the addition of the event. This configuration can be described as a set ($K$) of tuples $\langle te_s,te_d,\lambda,\kappa\rangle$:
\begin{itemize}
  \item the element $te_s$ is a triggering event that is to be observed by the capability. The set of observed events is given by the set $S$ ($te_s \in S$), which is a subset of the available reasoning events (\(S \in E\)).
  \item the element  $te_d$ is the corresponding event that is to be added to the agent reasoner when $te_s$ is observed. The set of actuated events ($D$) is also a subset of the available reasoning events (\(te_d \in D,D \in E\)). 
  \item the element $\lambda$ denotes the logical location of event additions. Events ($te_d$) can be placed in the currently active intention ($i$) or in a new intention ($\top; \lambda \in \{i,\top\}$).
  \item the optional element $\kappa$ is a boolean expression that defines when the event $te_d$ is applicable to be added to the agent configuration. This expression takes into account the current agent state ($ag,C,M,T$) and denotes the context that permits the introduction of the additional event ($te_d$). The expressiveness of these statements depends on the utilized agent platform. 
\end{itemize}

Three functions are introduced, which access this configuration, to simplify the semantics of agent state modifications. An auxiliary mapping function \(m(x)\) is assumed that maps triggering events to corresponding events ($m: S \longrightarrow D$). According to the configurations in $K$ this function returns the event(s) that are to be introduced ($m(te_s) = te_d$). The function $l: S,D \longrightarrow \lambda$ returns the target intention ($\lambda$) for an event mapping ($S,D$), i.e. two corresponding events ($te_s,te_d$). $\lambda$ can have two values and indicates either that the event is to be added to the current intention ($\lambda_{c}$) or that the event is to be included in a new intention ($\lambda_{n}$). In the following, we also assume the availability of a platform specific function $eval(te_s,te_d)$ that extracts the agent execution context, looks-up the corresponding condition statement ($\kappa$) and evaluates the statement, i.e. returns a boolean ($eval: S,D \longrightarrow \{true,false\}$). The functions $m$ and $l$ are prescribed by the agent programmer. The given mapping defines the set of observed events and their counterparts, which are to be induced. For each of the latter events, its is also specified whether it is to be placed in the current intention or in a new one. New intentions initially contain only the triggering event and this placement initializes another concurrent course of actionfor the agent.

Implementations of CECs are aware of the mapping function and dispatch the corresponding event in the surrounding agent. Afterwards the agent execution continues unaltered. Therefore, only the event selection rule \emph{SelEv} is supplemented with another rule (\emph{SelEvCEC}) that defines the the contribution of a CEC when events are selected that are in the set of observed events \(S\). If \(te \notin S\), the unaltered selection rules (SelEv1,2) is used (cf. Section \ref{bdi-so}). The inserting of the event ($te_d \in D$) that corresponds the observed event \(m(te)\) into the agent circumstance ($C'_E$) is guarded by the annotated condition ($\kappa$) that is evaluated for the currently handled event mapping ($eval(te,m(te))$). Events are inserted into the intention that is indicated by the function $l(te_s,te_d)$. The original event ($te$), which was selected for processing ($S_E(C_E) = \langle te,i\rangle$), is added to the temporal structure ($T'_{\epsilon}$) and subsequently processed by the reasoning cycle. This rule does not enforce immediate event processing. The additional event ($te_d$) is added to the events that will be subsequently processed by the agent but the operation of the interpreter decides how agent execution proceeds.
\begin{equation}\label{equation:the-equation}
	\textbf{SelEvCEC}\quad \frac{S_{E}(C_{E}) = \langle te,i\rangle \qquad (te \in S)}{\langle ag,C,M,T,SelEv\rangle \longrightarrow \langle ag,C',M,T',RelPl\rangle}
\end{equation}
where:
\[ 
C'_{E} = \left\{ 
\begin{array}{l l}
  C_{E} \cup \{\langle m(te),l(te,m(te) \rangle\} \backslash \{\langle te,i \rangle\} & if \, eval(te,m(te)) = true \\
  C_{E} \backslash \{\langle te,i \rangle\} & otherwise
\end{array}
\right.
\quad \quad \quad \quad \quad T'_{\epsilon} = \langle te,i\rangle
\]
\subsection{Prototype Realization}\label{impl}
The described mechanism has been implemented using the \emph{Jadex} reasoning engine\footnote{http://jadex.informatik.uni-hamburg.de}, that provides an execution environment for BDI-style agents on top of arbitrary distributed systems middleware \cite{Braubach2005b}. 
The Jadex platform allows implementations of a certain interface (jadex.run-time.ISystemEventListener), to be registered at individual agents. The \emph{Jadex Introspector} and \emph{Tracer} development tools, accompanying the Jadex distribution, use this mechanism to observe agent reasoning. Implementations of this interface can be registered for a specified set of reasoning events (defined in jadex.runtime.Systemevent) and an API allows to modify the BDI facilities of the surrounding agent via an API.

\subsection{Structuring Agents with Active, Context-Aware Modules}\label{cc_usage}
In terms of \emph{comprehensibility} and \emph{reusability}, it is advisable to cluster  distinguishable functionalities in modules. Capabilities contain libraries of BDI elements that can be referenced to reuse functionality. Besides, the goal-oriented and event-based processing of the reactive planning mechanism, can as well be exploited to handle functionalities that would be commonly understood as crosscutting concerns. For example, message based communication are encapsulated in \cite{Garcia2004} in an \emph{interaction} aspect while the modularization of roles in a negotiation protocol is the prime example for BDI-based capabilities \cite{Busetta2000} (cf. section \ref{bdi-se}). The proposed extension to the capability concept allows to automate the references to elements that are contained in capabilities. This facilitates information-hiding, as both the agent elements and the information when these are to be activated/modified are contained in a single entity. 


In addition, this approach facilitates the provision of \emph{additional}, quasi \emph{contributive}, processing of BDI reasoning events. The original handling of events leads to a series of subsequent events. For example the achievement of a goal may involve the successive achievement of subgoals. Besides this sequential execution path, which is controlled by the agent reasoner, the described mechanism allows to declare additional activities should be activated as well. These are caused by the activation of reasoning events (see equation \ref{equation:the-equation}). The authority of the reasoning mechanism, e.g. a BDI interpreter, is untouched and the contributive events integrate in the currently active or a in a concurrently executed intention. 
An example usage is the embedding of monitoring routines that decide the significance of agent state changes and communicate these to remote observers \cite{Sudeikat2006} as well as the embedding of assertion validations \cite{Sudeikat2006a}.

The here discussed \emph{activation} of agent modules reveals another perspective on the modularization of agents. Besides functional clusters (see Section \ref{bdi-se}), agents can be structured as composites of context-aware actors that decide locally when to provide the contained functionalities. Despite the outlined benefits, the presented mechanism reduces the \emph{traceability} of the agent actions. it makes it more difficult to trace the causes of specific agent actions. Since the additional events are handed over to the agent reasoner for subsequent processing, arbitrary agent events can be selected for inclusion. Subsequently, all events are similarly processed. Therefore, coefficient activities can be arbitrarily selected. However, the performance of the agent functionality is reduced by the coefficient inclusion of exhaustive processings. 

\subsection{Construction of Coordination Endpoints}
A requirement for the realization of the Decomas architecture is the ability to associate Endpoints with agents and enable these endpoints to observe, reason about, and influence the agent execution (see Section \ref{sect:architecture}). The discussed coefficiency mechanism allows to construct Coordination Endpoints as agent modules. Conceptually, this approach is attractive since the agent coordination is separated is represented as an orthogonal aspect. The uncommunicative overhead, imposed by remote endpoints is avoided. Using coefficient agent modules, a generic Endpoint model has been conceived that abstracts from the utilized agent platform.

\begin{figure}[htp]
\begin{center}
  \includegraphics[width=1.0\textwidth]{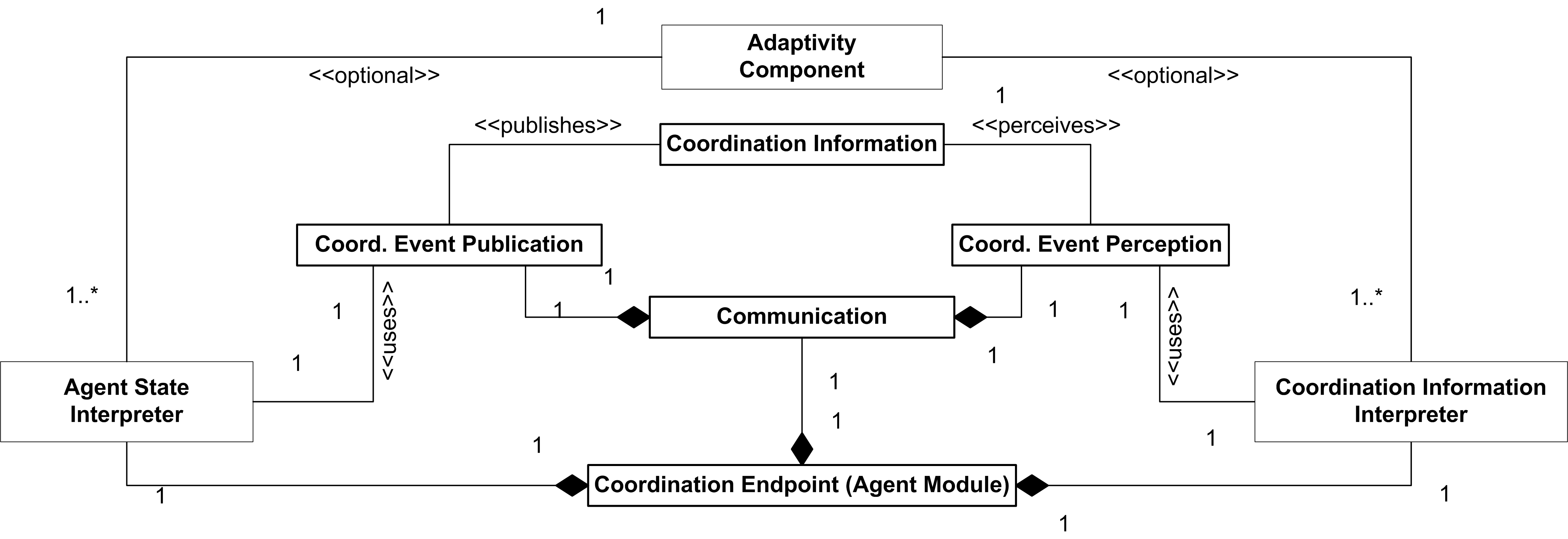}
  \caption{Excerpt from \cite{Sudeikat2009j} that illustrates the structure of Coordination Endpoints.}
  \label{fig:coordination-architecture-excerpt}
\end{center}
\end{figure}

This structure is illustrated in Figure \ref{fig:coordination-architecture-excerpt}. A \emph{Coordination Endpoint} is composed of three (sub-) modules. First, a \emph{Communication} module contains the abilities to \emph{publish} and \emph{perceive} events that are significant for the inter-agent coordination. These modules exchange specific data elements (\emph{Coordination Information}) and realize the Medium-specific information propagation. Endpoints contain also two interpreter elements. The \emph{Coordination Information Interpreter} is responsible to process the perceived information and decides, based on a declarative configuration of the enacted coordination process \cite{Sudeikat2009}, the appropriate modifications of the host agent. This is a conventional agent module except that it affects the injection of modifications in the surrounding agent (see Section \ref{cc-so-detail}). The \emph{Agent State Interpreter} is a coefficient module that registers for the observation of the surrounding host agent. The events of interest are given by a declarative model of the desired inter-agent coordination \cite{Sudeikat2009} that also contains the conditions and constraints that control the publication of agent-internal events and data. Using \emph{coefficiency}, the observation and affection of publications is realized as an autonomous background process inside agents. 
\section{Shoaling Glassfishes: Decentralized (Web) Service Management}\label{sect:shoaling_glassfishes}
Distributed software systems imply an administrative overhead that originates in the  manual maintenance of computational infrastructure, e.g. the provision of server installation and server deployments in \emph{Service Oriented Architectures}. The reduction of this overhead in dynamic environments, where  request loads and resource usages fluctuate, is a prominent research topic \cite{Kephart2008}. Often it is necessary to augment existing middlewares with adaptive mechanisms \cite{Garlan2004}. 
 
\begin{figure}[htp]
\begin{center}
  \includegraphics[width=0.55\textwidth]{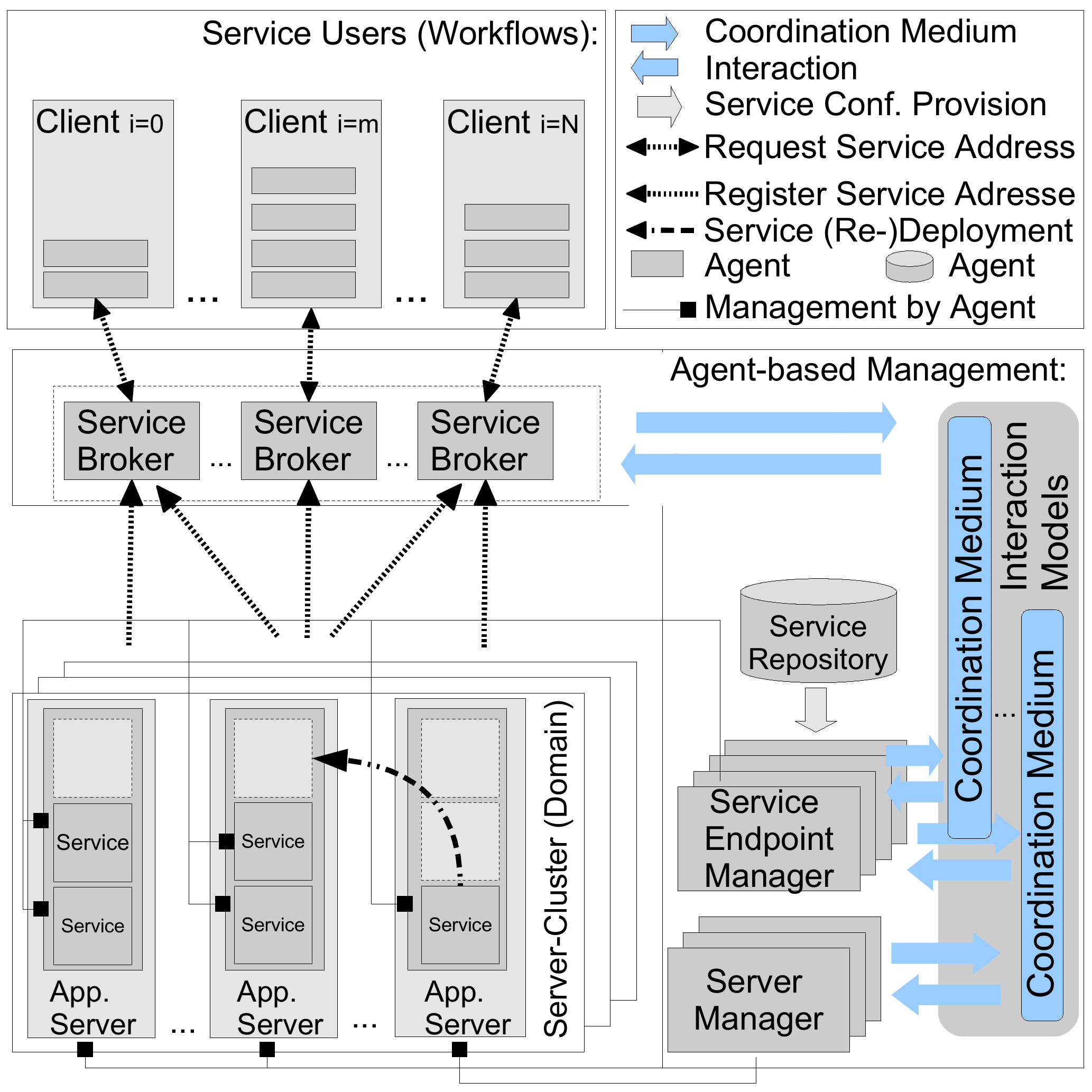}
  \caption{Decentralized, agent-based service management framework.}
  \label{fig:architecture}
\end{center}
\end{figure}

Here, we outline the development of a \emph{decentralized}, \emph{agent-based} management framework for the maintenance of distributed software infrastructures.  
Figure \ref{fig:architecture} illustrates the conceptual architecture. \emph{Services} are deployed on application server instances (\emph{App. Server}). These servers reside on different \emph{Server-Clusters}, i.e. are distributed in different computing / data centers. Service endpoints and application servers are managed by remote software agents (\emph{Agent-based Management}). These agents monitor and manipulate the configurations of services and servers. The management by agents is realized with the SUN \emph{Appserver Management EXtensions}\footnote{https://glassfish.dev.java.net/javaee5/amx/index.html} (AMX), a generic API to control J2EE Application Server configurations. Managers of application servers are bound to specific installations, while service endpoints are free to reallocate to different servers and change their service offer. Agents are equipped with plans for the deployment and undeployment of services. Unknown  of service configurations can be fetched from remotely accessible \emph{repositories}. Service consumers (cf. figure \ref{fig:architecture}, top-left) invoke web services. The dynamics of service deployments and server utilizations are hidden by \emph{Service Brokers}. These maintain local registries of the physical locations of service deployments. Therefore, clients can retrieve the addresses of the current deployments from the static locations of the Brokers. In addition, the Brokers load-balance the service utilizations.

This prototype architecture has been realized with the \emph{Jadex} agent framework (cf. Section \ref{impl}). It prepares the agent-based management of service infrastructures as agents 
are capable to administrate services and servers, i.e. to deploy / undeploy services. The server management has been tested with the freely available \emph{Glassfish}\footnote{Version 2 (ur2-b04),  https://glassfish.dev.java.net/} application server. The actual coordination logic is supplemented with the systemic programming model that is discussed in Section \ref{sect:architecture}. This management concerns two aspects of the dynamic deployment of (web) services.  First, the allocation of physical servers is balanced to maintain averaged utilization levels. Servers are associated with preferential workload-levels. Based on the communication of available capacities, services are moved to ensure that all servers are in their preferred operational condition. Secondly, the adjustment of static service deployments is automated to meet dynamically changing service workloads. The deployments of highly-demanded services are reinforced and less-demanded services are reduced.

\subsection{Server Utilization Management}\label{sect:integrating_servier_utilization_management}
The aim of the utilization management is to maintain a preferential number of service deployments on servers. The rational is, that servers are dimensioned for a specific utilization. Therefore, the maintenance of several  underutilized servers is a waste of resources, e.g. energy \cite{Kephart2008}, when the services could be handled by a single server that is properly utilized. The decentralized management is based on the publication of capacities by underutilized servers and services are attracted to servers that are almost well utilized. 

\begin{figure}[htp]
\begin{center}
\includegraphics[width=0.92\textwidth]{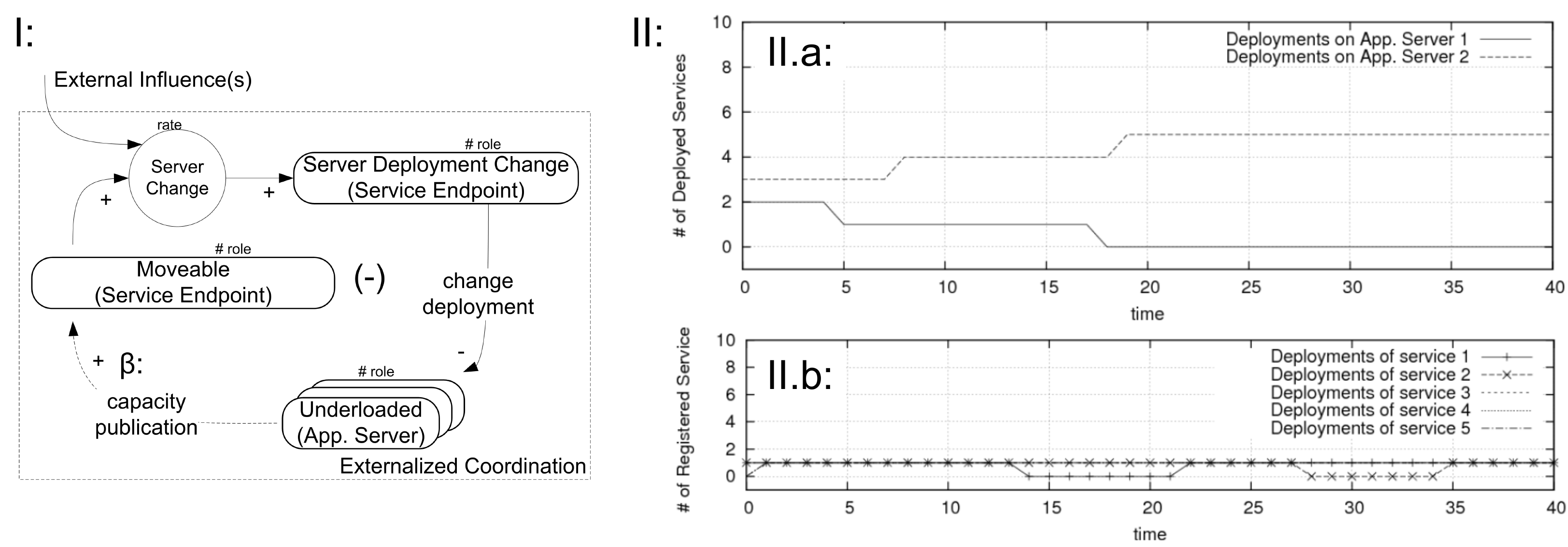}
  \caption{Server utilization management: The dynamics of the management (I) and a simulation snapshot (II). The simulation results relate the deployments per server (II.a) to all service registrations (II.b).}
  \label{fig:case_study_2}
\end{center}
\end{figure}

The dynamics of the adaptive placement of services is illustrated in Figure \ref{fig:case_study_2} (I). The variable \emph{Underloaded} describes the number of servers that are below their intended utilization. This utilization is maintained by a balancing feedback loop \emph{(-)}. 
A Coordination Endpoint determines whether the server is underutilized or not (a logical condition, defined on the set of agent beliefs). The Endpoints within underloaded servers publish the availability of capacities ($\beta$). These publications propagate in a Coordination Medium to the Coordination Endpoints in Service Endpoint agents (\emph{Moveable}). These decide whether to change the service deployment, i.e. move to another server, or not. Movements affect the system-wide rate of (re-)deploying agents (\emph{Server Change}) and consequently influences the number of (re-)deployed services. Since these deployment actions were caused by the availability of resources, the number of underloaded servers decreases. 

Figure \ref{fig:case_study_2} (II) shows simulation results for the described coordinating process. Two application servers are initialized with slightly different workloads. Servers are configured to host up to 5 services. The availability of server capacity gradually spreads in the system and services are (re-)deployed. Figure \ref{fig:case_study_2} (A) denotes the movement of services that is carried out be their un- and (re-)deployment. 
\subsection{Balancing Service Deployments}\label{sect:integrating_service_balancing}
The adaptive reinforcement of service deployments, balanced with fluctuating request loads, is based on the publications of demand changes by Broker agents. Using the coordination architecture, the measurement and interpretation of demand changes is encapsulated in Endpoint modules. These publish significant changes and upon the reception of these information, the Endpoints of Service agents decide whether to adjust the local deployment or not.

\begin{figure}[htp]
\begin{center}
  \includegraphics[width=1.0\textwidth]{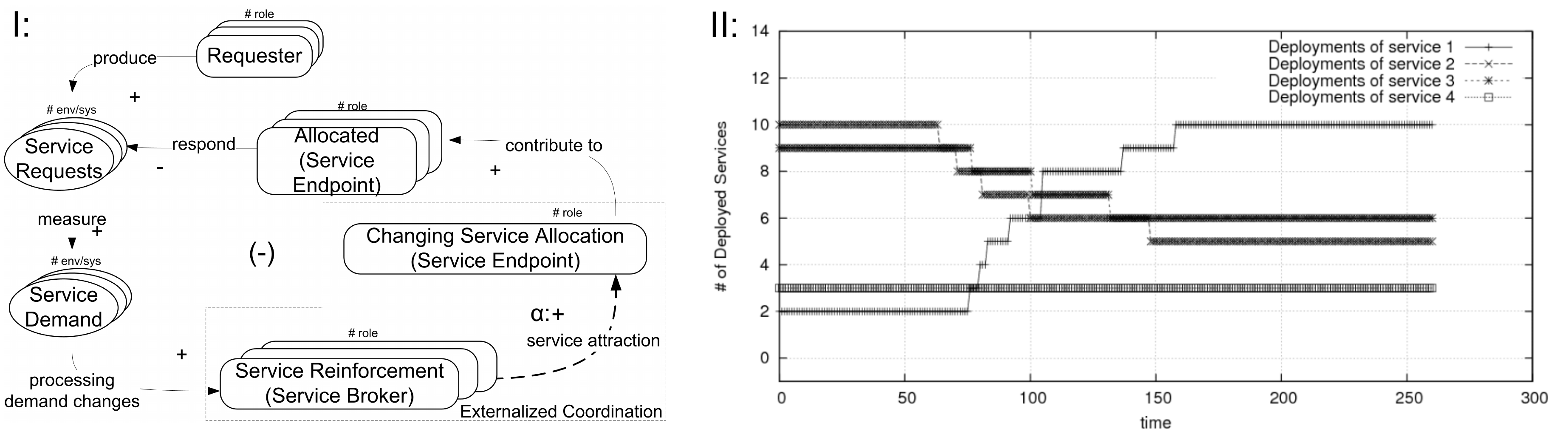}
  \caption{J2EE (Web-)Service management: Balancing Service Workload.}
  \label{fig:case_study_1}
\end{center}
\end{figure}

The dynamics of this adaptive process is illustrated in Figure \ref{fig:case_study_1} (I) 
\emph{Requesters} increase, with a fluctuating rate, the number of \emph{Service Requests} that the system has to work off. The amount of requests causally influences the measured \emph{Service Demand}. The \emph{Allocated} variable denotes the number of agents that offer services. 
Service Brokers are supplemented with the ability to publish ($\alpha$) significant changes in request workloads (\emph{Service Reinforcement}). The publications affect that agents consider an adjustment of their current service offer (\emph{Changing Service Allocation}). The additional behaviors, required to participate in this process, are implemented in Coordination Endpoints. Within \emph{Service Brokers}, these decide the significance of workload changes and in \emph{Service Endpoints}, these decide whether to change deployments or not.

Figure \ref{fig:case_study_1} (II) shows simulation results for 10 virtual application servers (domains) that run on a Glassfish installation. Each of them is configured to host a maximum of 5 web service implementations. An additional constraint is that every service type is only deployed once on every domain. Initialized with an arbitrary configuration of service deployments, the system is exposed to a sudden workload of service type 1. The publication of this demand change enforces that Service Endpoints locally adjust and switch their deployments. Details on the integration of this coordination model and the declaration of agent-internal data to be communicated, including code fragments, can be found in \cite{Sudeikat2009c,Sudeikat2009}.

\section{Conclusions}\label{sect:conclusions}
In this paper, we presented an architecture for the integration of decentralized, self-organizing coordination in MAS. This architecture provides a middleware layer that contains coordination mechanisms and automates their invocation. A key design concern is that the architecture can be integrated in established agent architectures and that coordination can be supplemented to functional MAS. This allows that self-organization can be supplemented to conventionally developed MAS. The accomplishment of this design objective requires that invocations can be supplemented without affecting the structure of the original, i.e. not self-organizing, MAS. This has been approached with the concept of \emph{activated} agent modules. Their generic structure and operating principle is discussed. Their implementation is outlined and formalized for a particular agent architecture. The utilization of the coordination framework is exemplified for the management of service oriented computing infrastructures. Future work comprises the programming language techniques that are used to control the execution of Media and Endpoint elements. Ongoing work concerns the revision of a declarative configuration approach \cite{Sudeikat2009} to integrate the prescriptions of self-organizing processes in MAS development frameworks.


\section*{Acknowledgment}
We cordially thank Rafael H. Bordini\footnote{R.Bordini@inf.ufrgs.br
} from the \emph{Institute of Informatics} at the Federal University of Rio Grande do Sul for discussing and reviewing the formalization of the endpoint operation principle that is discussed in Section \ref{op-sem}. We also thank the \emph{Distributed Systems and Information Systems} (VSIS) group at Hamburg University, particularly W. Lamersdorf, L. Braubach, A. Pokahr, and A. Vilenica for helpful discussion. The SodekoVS-project is funded by the \emph{Deutsche Forschungsgemeinschaft} (DFG)\footnote{http://www.dfg.de}.
\bibliographystyle{eptcs} 

\end{document}